\documentclass[twocolumn,11pt]{article}

\usepackage{amssymb,amsmath}
\usepackage{graphicx}

\usepackage{hyperref}
\usepackage{ragged2e}
\usepackage[USenglish]{babel}
\usepackage{hyphsubst}

\begin{document}

\author{Guzm\'an V.}

\twocolumn[

\centering
{\LARGE A geometric description of blackbody-like systems in thermodynamic equilibrium \\[0.5em]
\large Guzm\'an R. Victor A. }\\
Departamento de F\'{\i}sica - FaCyT, Universidad de Carabobo, Valencia 2001, Venezuela\vspace*{0.5cm}

\centering{\large \textbf{Abstract}}\\
\justifying Riemannian and contact geometry formalisms are used to study the fundamental equation of electromagnetic radiation-like systems, obeying a Stefan-Boltzmann's-like law. The vanishing of metric determinant is used for classifying what kind of systems can not represent a possible generalization of blackbody-like systems. In addition, thermodynamic curvature scalar $\mathcal{R}$ is evaluated for a thermodynamic metric, giving $\mathcal{R}=0$, which validates the non-interaction hypothesis stating that the scalar curvature vanishes for non-interacting systems.
\vspace*{1cm}
]






\section{Introduction}

In a series of 5 papers, Weinhold studied the implicit relationship between the empirical laws of thermodynamics and the axioms of an abstract metric space, constructing some vectors and their operations in a similar way as in quantum mechanics for quantum states \cite{Weinhold-1975.I,Weinhold-1975.II,Weinhold-1975.III,Weinhold-1975.IV,Weinhold-1976.V}. He used this for multiple reasons, one of them, for the creation of a Riemannian metric which was given by the Hessian of the fundamental equation as given by Gibbs \cite{Gibbs-1873}, this work let Ruppeiner to construct another Riemannian metric by different considerations that is conformally equivalent to Weinhold's metric \cite{Ruppeiner-1979}. In this research Ruppeiner introduced an important hypothesis saying that, the metric can be interpreted as a measure of the thermodynamic interaction, in analogy with general relativity assumption. These works open the idea for a Riemannian geometric representation of thermodynamics.

Another important set of contributions were given by Mr\"uga{\l}a, who constructed the thermodynamic phase space and showed that thermodynamics has an implicit contact structure \cite{Mrugala-1978}. Then he reformulated the whole thermodynamic theory in terms of differential geometry for contact manifolds. Finally Quevedo presented a program that he called \emph{geometrothermodynamics} \cite{Quevedo-2007} where he uses all of this contributions to construct a thermodynamic metric which is invariant under Legendre transformations, then he uses this metric to study how curvature is related to phase transitions for different systems \cite{Quevedo-2011,Aviles-2012,Bravetti-2014}.

In view of all of this contributions, a book was recently published that summarizes some of the geometric approaches given to the thermodynamic theory and other ones that will not be considered for this research\cite{Badescu-2016}. In this research the approach will be to study the traditional problem of blackbodies, and a generalization on these kind of systems by means of a Neo-Gibbsian approach \cite{Callen-1985}, and the recent developments on the field.

This paper is organized as follow: in section 2, the problem of blackbody-like systems will be discussed from the thermodynamic point of view, in the next section a Riemannian metric will be constructed in the thermodynamic phase space of equilibrium states, and the metric's determinant will be analized for those cases where it vanishes, then in section 4, metric's curvature will be calculated and interpreted in terms of an interaction hypothesis, and finally in section 5 these methods are used to interpret the fundamental equation for these systems.

\section{Graybodies: Neo-Gibbsian Picture}

The empirical equations that characterize thermodynamically the blackbody are:

\begin{subequations}
\begin{equation}\label{SB-BBE}
U=\left(\frac{4\sigma}{c}\right)VT^{4}
\end{equation}
\begin{equation}\label{P-ER}
P=\frac{U}{3V},
\end{equation}
\end{subequations}
where $S,U$ and $V$ are entropy, internal energy and system's volume respectively and $\sigma$ and $c$ are Stefan-Boltzmann's constant and light's speed respectively. This two equations can be used to construct a fundamental equation in the entropic representation for electromagnetic radiation\cite{Callen-1985}:

\begin{equation}\label{Fun-Eq-ER-1}
S(U,V)=\frac{4}{3}\left(\frac{4\sigma}{c}\right)^{1/4}U^{3/4}V^{1/4}.
\end{equation}

In the equation above, particle's number $N$ doesn't appear, because the two empirical equations that characterize electromagnetic radiation don't have a dependence on $N$, so in this system it is not possible to think about conserved particles to be counted by a parameter $N$.

The blackbody system can be generalized by considering the equation for a graybody:
\begin{equation}
U=\varepsilon\left(\frac{4\sigma}{c}\right)VT^{4},
\end{equation}
where $\varepsilon\in(0,1]$, the value $\varepsilon=1$ brings back the equation for a blackbody \ref{SB-BBE}, and the value $0$ can not account for a physical system because it would mean that internal energy could be independent of system's state and equals to $0$. Taking this into account it can be obtained a new fundamental entropic equation for graybodies:

\begin{equation}\label{Fund-Eq-ER-2}
S=\frac{4}{3}\left(\frac{4\sigma\varepsilon}{c}\right)^{1/4}U^{3/4}V^{1/4}.
\end{equation}
An additional generalization can be done by considering that the equation \ref{Fund-Eq-ER-2} is a particular case of:

\begin{equation}\label{Fun-Eq-ER}
S=bU^{a}V^{1-a},
\end{equation}
with $b=(4/3)(4\sigma\varepsilon/c)^{1/4}$ and $a=3/4$. This fundamental entropic equation accounts for systems with \emph{empirical equations} of state given by:
\begin{subequations}
\begin{equation}\label{EES-U}
U=a^{(1+a)/(1-a)}b^{1/(1-a)}VT^{(1+a)/(1-a)}
\end{equation}
\begin{equation}\label{EES-P}
P=\left(\frac{1-a}{a}\right)\frac{U}{V}.
\end{equation}
\end{subequations}

This kind of systems described by \ref{EES-U},\ref{EES-P} will be analized in terms of a Riemannian metric.

\section{Thermodynamic Metric Restrictions}

The smooth $(2(2)+1)-$dimensional $\mathcal{T}$ thermodynamic phase space is a contact manifold which admits a local coordinate representation given by Darboux's theorem:

\begin{equation}
\Theta=d\Phi-I_{1}dE_{1}-I_{2}dE_{2},
\end{equation}
where $\Phi$ is the fundamental equation, $I_{i}$ and $E_{i}$ are the intensive and extensive parameters for $\Phi$ being in entropic or energetic representation. The pair $(\mathcal{T},\Theta)$ denotes a contact manifold where $\Theta$ satisfies a non-degeneracity condition: 

\begin{equation}
\Theta\wedge(d\Theta)^{2}\neq 0.
\end{equation}

It is possible to find the smooth thermodynamic equilibrium phase space of the system $\mathcal{E}$ as the space spanned by the extensive parameters $E_{1},E_{2}$, and this is possible by means of the embedding mapping $\varphi: \mathcal{E}\rightarrow\mathcal{T}$, then  $\mathcal{E}$ is an integral manifold given by the $\Theta$ field if:

\begin{equation}
\varphi^{*}(\Theta)=\theta=0,
\end{equation}
where $\varphi^{*}$ represents the pullback. For a 2-dimensional system given in an entropic representation:

\begin{equation}
\theta_{S}=dS-\frac{1}{T}dU-\frac{P}{T}dV=0.
\end{equation}

This last equation is another way for writing the first law of thermodynamics for reversible processes. For the system \ref{Fun-Eq-ER}:

\begin{equation}
\theta_{S}=dS-ab\frac{U^{a-1}}{V^{a-1}}dU+(1-a)b\frac{U^{a}}{V^{a}}dV=0.
\end{equation}

Now as given in the geometrothermodynamics program \cite{Aviles-2012}. A differential thermodynamic length invariant under Legendre transformations is given by:
\begin{align}
&dL_{S}^{2}=\left(dS-\frac{1}{T}dU-\frac{P}{T}dV\right)^{2}+\cdots \nonumber\\
&\cdots+\Lambda\left[U\left(\frac{1}{T}\right)dUd\left(\frac{1}{T}\right)+V\left(\frac{P}{T}\right)dVd\left(\frac{P}{T}\right)\right].
\end{align}
Calculating $dL_{S}^{2}$'s pullback, $\varphi^{*}(dL_{S}^{2})=dl_{S}^{2}$:
\begin{equation}
dl_{S}^{2}=\Lambda\left(US_{U}dUdS_{U}+VS_{V}dVdS_{V}\right),
\end{equation}
this can be re-written as:
\begin{align}
dl_{S}^{2}=&\Lambda US_{U}S_{UU}dU^{2}+\Lambda S_{UV}dUdV+\cdots\nonumber\\
&\cdots+\Lambda VS_{V}S_{VV}dV^{2}.
\end{align}
The associated metric can be written as:
\begin{equation}
g_{S}=\left( \begin{array}{ccc}
\Lambda US_{U}S_{UU} & \frac{1}{2}\Lambda SS_{UV}  \\
\frac{1}{2}\Lambda SS_{UV} & \Lambda V S_{V}S_{VV}   \end{array} \right)
\end{equation}
where $S_{U}=aS/U$, $S_{V}=(1-a)S/V$, $S_{UU}=-a(1-a)S/U^{2}$, $S_{UV}=a(1-a)S/UV$ and $S_{VV}=-a(1-a)S/V^{2}$, using this:
\begin{equation}\label{gs-metric}
g_{S}=\left( \begin{array}{ccc}
-\Lambda a^{2}(1-a)\frac{S^{2}}{U^{2}} & \frac{1}{2}\Lambda a(1-a)\frac{S^{2}}{UV} \\
\frac{1}{2}\Lambda a(1-a)\frac{S^{2}}{UV} & -\Lambda a^{2}(1-a)\frac{S^{2}}{V^{2}}   \end{array} \right)
\end{equation}
it can be seen that:
\begin{equation}\label{det-gS}
\det g_{S}=\Lambda^{2}a^{2}b^{4}(1-a)^{2}\left(a^{2}-\frac{1}{4}\right)\left(\frac{U}{V}\right)^{4a-2}.
\end{equation}

Now the cases for which $\det g_{S}\neq 0$ will be analyzed, obtaining:
\begin{subequations}
\begin{equation}
\Lambda\neq 0
\end{equation}
\begin{equation}\label{a-0,1}
a\neq \left\lbrace-\frac{1}{2},0,\frac{1}{2},1\right\rbrace
\end{equation}
\begin{equation}
b\neq 0
\end{equation}
\end{subequations}
the case for $\Lambda=0$ is the trivial one because $dl^{2}=0,\;\forall\; (U,V,a,b)$ so all different states could be represented by the same one. This case can't be seen as a possible system. The other cases have a different interpretation. For $a\neq 0$, it can be seen from \ref{EES-P} that if $a\rightarrow 0$ any small variation of internal energy $\delta U$ or volume $\delta V$ would result on $P\rightarrow \infty$, that is, in any local chart on the manifold $\mathcal{E}$, the tangent space $T_{q}\mathcal{E}$ could not be defined because the element $S_{V}$ can not be properly defined on the smooth manifold $\mathcal{E}$. So this case couldn't represent a physical system. In addition for $P$ being possitive: 
\begin{equation}\label{a-rest-1}
0\leq a\leq 1,
\end{equation}
so restriction $a\neq -1/2$ is considered by \ref{a-rest-1}, in addition let's recall that $a\neq 0$, means that $0<a\leq 1$. The implications for the case $a=1$ can be seen for example in \ref{EES-U}, for $a=1$ energy couldn't be defined, so $0<a<1$. These restrictions are obtained by considering $\det g_{S}\neq 0$. The case for $b=0$ can be seen in \ref{EES-U}, for this case energy couldn't be defined. 

Finally the last restriction $a\neq1/2$ is interpreted in a different way, according to \ref{det-gS}, $a=1/2$ splits the region of possible values for $a$ and $b$, depending on $\det g_{S}$ sign. It can be seen that for $\lim_{a\rightarrow 1/2}\det g_{S}$, system's determinant becomes independent of system's state $(U,V)$.

\begin{figure}
\centering
\includegraphics[scale=0.38]{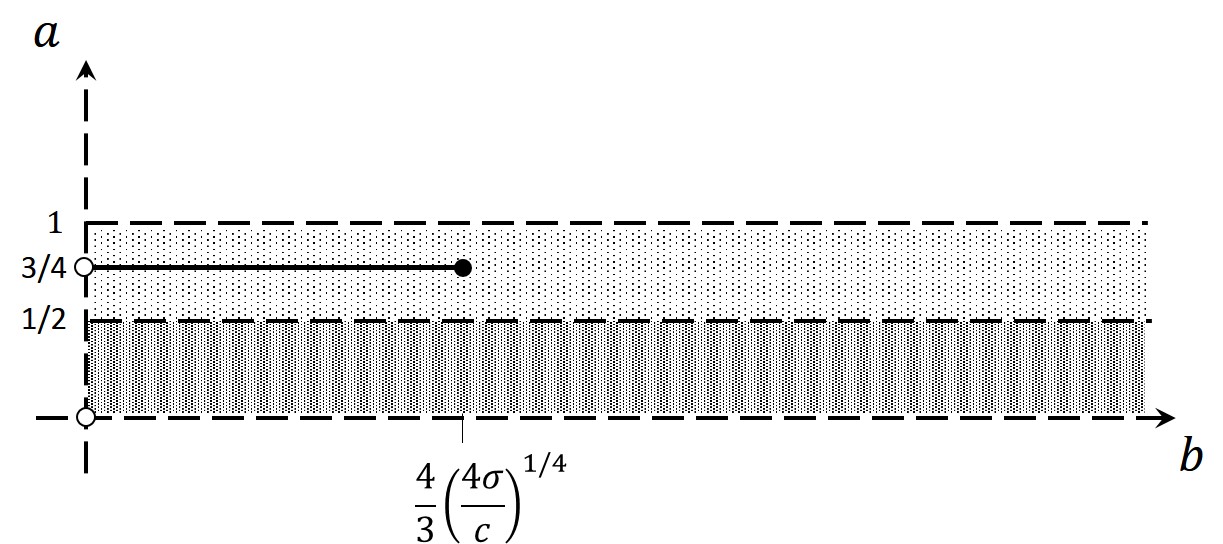}
\caption{Region of possible physical systems described by the fundamental equation \ref{Fun-Eq-ER}. The solid line at $a=3/4$ can be interpreted as thermodynamic graybody-like systems, the final dot at $b=(4/3)(4\sigma/c)^{1/4}$ represents the blackbody and the dash line at $a=1/2$ splits this region in two, according to $\det g_{S}$ sign.}\label{Restrict-ab}
\end{figure}

According to determinant's sign, the value for graybodies $a=3/4$, are found in the upper region for which $\det g_{S}>0$ as can be seen in figure \ref{Restrict-ab}. 

An interesting result showing $a=1/2$ as an \emph{special} value, can be obtained by considering the explicit relationship between the fundamental equation for entropy $S$ and its response function $C_{V}$:
\begin{equation}
C_{V}=\left(\frac{a}{1-a}\right)S,
\end{equation}
so $C_{V}\leq S$, obtaining $C_{V}=S$ only for $a=1/2$. So this splitting behaviour observed for $a=1/2$, can be seen as the only one system in which its entropy have no differences with its response function. Remembering that $S$ is the fundamental equation for these systems, this specific value $a=1/2$ implies that this response function have all of the system's information.

\section{Curvature and Interaction}

Let $\mathcal{R}$ be the curvature scalar associated to a thermodynamic metric $g$ \cite{Quevedo-2007} then:

\begin{align}
\mathcal{R}=&-\frac{1}{\sqrt{\det(g)}}\Bigg[\frac{\partial}{\partial V}\left(\frac{g_{11,V}-g_{12,U}}{\sqrt{det(g)}} \right)+\cdots\nonumber\\
\cdots+&\frac{\partial}{\partial U}\left(\frac{g_{22,U}-g_{12,V}}{\sqrt{det(g)}} \right)\Bigg]-\frac{det(H)}{2det(g)^{2}},
\end{align}
being the metric $g$ in the equation above given by \ref{gs-metric}, where $g_{ij}$ represents the elements of the matrix $g_{ij,x}=\partial g_{ij} \backslash \partial x$ and $H$ is:

\begin{equation}
H=\left( \begin{array}{ccc}
g_{11} & g_{12} & g_{22}  \\
\frac{\partial g_{11}}{\partial U} & \frac{\partial g_{12}}{\partial U} & \frac{\partial g_{22}}{\partial U} \\
\frac{\partial g_{11}}{\partial V}  & \frac{\partial g_{12}}{\partial V} & \frac{\partial g_{22}}{\partial V}   \end{array} \right).
\end{equation}

The scalar curvature for the two-dimensional metric system will be \ref{Fun-Eq-ER}:

\begin{equation}\label{R-0}
\mathcal{R}=0,\qquad \forall (U,V)
\end{equation}
and also for all $a$ and $b$.

Now this result can be interpreted by means of the interaction hypothesis \cite{Ruppeiner-1979}. This hypothesis states that the curvature scalar may be interpreted as: 

\begin{itemize}
\item $\mathcal{R}$ zeros implies no-interaction
\item $\mathcal{R}$ poles implies phase transitions \cite{Quevedo-2011},\cite{Aviles-2012}.
\end{itemize}

This means that for any system characterized by a fundamental equation as \ref{Fun-Eq-ER}, there is no internal interaction.
\section{Conclusions}
In this work, the formalism of recent geometric approaches was applied, for interpreting the restrictions for a generalization to thermodynamic systems behaving like graybodies. First, the fundamental equation for this kind of systems was constructed and according to the selected metric \cite{Aviles-2012}, it was found that these systems have a flat equilibrium manifold as can be seen by $\mathcal{R}=0$, and this is interpreted as a non-interacting system. This result, does not depend on the particular state of the system $(U,V)$ or the particular generalized system $a,b$.

It was also discussed the cases for which $\det g_{S}=0$. Obtaining that these values can be seen as restrictions to the fundamental equation \ref{Fun-Eq-ER}, in addition it turned out to be that $a=1/2$, obtained as one of these restrictions, represents the only one value for which determinant sign is changed. This investigation shows that vanishing of metric's determinant have physical interesting implications, and that they can be understood as restrictions for a possible generalization of blackbody-like systems.

\bibliographystyle{unsrt}

\end{document}